\documentclass[english]{article}
\usepackage[T1]{fontenc}
\usepackage[latin9]{inputenc}
\usepackage{amssymb}
\usepackage{esint}
\usepackage{babel}
\begin{document}

\title{\textbf{Tunnelling mechanism in non-commutative space with generalized
uncertainty principle and Bohr-like black hole}}
\maketitle
\begin{center}
\textbf{\large{}Sourav Haldar$^{1}$, Christian Corda$^{2}$, Subenoy
Chakraborty$^{3}$}
\par\end{center}{\large \par}

\begin{center}
\textbf{\large{}$^{1}$Department of Mathematics, Jadavpur University,
Kolkata-700032, India. E-mail address: sourav.math.ju@gmail.com}
\par\end{center}{\large \par}

\begin{center}
\textbf{\large{}$^{2}$ Research Institute for Astronomy and Astrophysics
of Maragha (RIAAM), P.O. Box 55134-441, Maragha, Iran. E-mail address:
cordac.galilei@gmail.com}
\par\end{center}{\large \par}

\begin{center}
\textbf{\large{}$^{3}$ Department of Mathematics, Jadavpur University,
Kolkata-700032, West Bengal, India. E-mail address: schakraborty.math@gmail.com}
\par\end{center}{\large \par}
\begin{abstract}
The paper deals with non-thermal radiation spectrum by tunnelling
mechanism with correction due to the generalized uncertainty principle
(GUP) in the background of non-commutative geometry. Considering the
reformulation of the tunnelling mechanism by Banerjee and Majhi, the
Hawking radiation spectrum is evaluated through the density matrix
for the outgoing modes. The GUP corrected effective temperature and
the corresponding GUP corrected effective metric in non-commutative
geometry are determined using Hawking's periodicity arguments. Thus,
we obtain further corrections to the non-strictly thermal black hole
(BH) radiation spectrum which give new final distributions.

Then, we show that the GUP and the non-commutative geometry modify
the Bohr-like BH recently discussed in a series of papers in the literature.
In particular, we find the intriguing result that the famous law of
Bekenstein on the area quantization is affected neither by non-commutative
geometry nor by the GUP. This is a clear indication of the universality
of Bekenstein's result. In addition, we find that both the Bekentsein-Hawking
entropy and the total BH entropy to third order approximation are
still functions of the BH quantum level.
\end{abstract}
\begin{quote}

\end{quote}

\section{Introduction}

Hawking radiation \cite{key-1} in the tunnelling mechanism {[}2 -
11{]} is an elegant way to approach the particle creation caused by
the vacuum fluctuations near the BH horizon. The virtual particle
pair can be created either just inside the horizon or just outside
the horizon. For both the possibilities the negative energy particle
is absorbed by the BH, resulting a loss of the mass of the BH and
the positive energy particle moves toward infinity, causing the subsequent
emission of Hawking radiation. Considering contributions beyond the
semi-classical approximation in the tunnelling process, Parikh and
Wilczek \cite{key-2,key-3} formulated the non-thermal spectrum of
the radiation from BH which leads to interesting approaches \cite{key-13,key-14}
to resolve the information loss paradox of BH evaporation \cite{key-12}.
Also subsequently, by a novel formulation of the tunnelling formalism,
Banerjee and Majhi \cite{key-7}, directly derived the black body
spectrum for both bosons and fermions from a BH with standard Hawking
temperature. The analysis in \cite{key-7} was improved by one of
us, C. Corda \cite{key-15}, who found as final result a non-strictly
black body spectrum in agreement with the emission probability in
\cite{key-2,key-3}. This non-thermal spectrum is deeply interrelated
to the underlying quantum gravity theory. As a result, the particle
emission can be interpreted as a quantum transition of frequency $\omega$
between two discrete states \cite{key-14,key-19,key-20}. Thus, the
particle itself generates a tunnel through the BH horizon \cite{key-3,key-15,key-19,key-20}
having finite size. This solves a problem of the thermal approximation,
namely in that case the tunnelling points have zero separation and
hence there is no clear trajectory as there is no barrier \cite{key-3,key-15,key-19,key-20}.
Other aspects of tunnelling mechanism have been discussed in \cite{key-42,key-43}.

In this paper we analyse the corrections to the non-thermal spectrum
of Parikh and Wilczek due to the non-commutative geometry and the
GUP. It is shown that such corrections modify the Bohr-like BH recently
discussed in a series of papers in the literature {[}13,18-22{]}.
An important result will be that the famous law of Bekenstein on the
area quantization \cite{key-35} is affected neither by non-commutative
geometry nor by the GUP. We can consider this as a clear indication
of the universality of Bekenstein's result. We also find that both
the Bekentsein-Hawking entropy and the total BH entropy to third order
approximation are still functions of the BH quantum level.

The physical reasons that motivate this work are two. On one hand,
the background of non-commutative geometry is important because it
can cure, in principle, some of the usual problems encountered in
the description of the terminal phase of BH evaporation \cite{key-16}.
On the other hand, the GUP suggests a fundamental and discrete granular
structure of space with important implications in quantum gravity
\cite{key-21,key-22,key-23}. For the sake of completeness, we take
the chance to signal some recent important approach where GUP importance
in BH physics is emphasized {[}44 - 49{]}. 

\section{Basic equations for non-commutative space}

In non-commutative space the usual definition of mass density in the
form of Dirac delta function does not hold due to position-position
uncertainty relation. A particle mass, instead of being exactly localized
at the point is diffused in a region of linear size $\sqrt{\delta}$.
The intrinsic uncertainty can be expressed as the co-ordinate commutator:
$[x^{\mu},x^{\nu}]=i\,\theta^{\mu\nu}$, where the antisymmetric matrix
$\theta^{\mu\nu}$ characterizes the fundamental cell discretization
of space-time, analogous to the discretization of the phase-space
by the Planck constant $\hbar$. It should be noted that the effect
of non-commutativity is not visible at presently accessible energies,
\emph{i.e.} $\sqrt{\delta}<10^{-16}$ cm. In particular, one should
take into account the non-commutativity effect at distance $r\simeq\sqrt{\delta}$
. As a result, there will be density of energy and momentum. Thus,
in non-commutative geometry the mass density is described by a Gaussian
distribution of minimal width $\sqrt{\delta}$ as (we work in Planck
units \emph{i.e.} $G=c=K_{B}=\hbar=\frac{1}{4\pi G}=1$ in the following)
\cite{key-16,key-17,key-18} 
\begin{equation}
\rho_{\delta}(r)=\frac{M}{(4\pi\delta)^{^{3}/_{2}}}e^{-\frac{r^{2}}{4\delta}}\label{eq: 1}
\end{equation}
Here the non-commutative parameter $\delta$ is a small ($\sim$ unity
in Planck units) positive number. Hence the mass of the analogous
Schwarzschild BH is given by \cite{key-18} 
\begin{equation}
m_{\delta}(r)=\int_{0}^{r}4\pi(r')^{2}\,\rho_{\delta}(r')\,dr'=\frac{2M}{\sqrt{\pi}}\,\gamma\left(\frac{3}{2},\frac{r^{2}}{4\delta}\right),\label{eq: 2}
\end{equation}
where the (lower) incomplete gamma function is defined as \cite{key-18}
\begin{equation}
\gamma(a,x)=\int_{0}^{x}t^{a-1}e^{-t}\,dt.\label{eq: 3}
\end{equation}
Thus, in the limit $\delta\rightarrow0$, \emph{i.e.} $x\rightarrow\infty$,
we have the usual gamma function and commutative geometry is recovered.
Hence, the usual Schwarzschild metric, i.e. \cite{key-15} 
\begin{equation}
ds^{2}=-(1-\frac{2M}{r})dt^{2}+\frac{dr^{2}}{1-\frac{2M}{r}}+r^{2}(\sin^{2}\theta d\varphi^{2}+d\theta^{2}),\label{eq: 4}
\end{equation}
takes the non-commutative form as \cite{key-18} 
\begin{equation}
ds^{2}=-\left(1-\frac{2m_{\delta}}{r}\right)dt^{2}+\left(1-\frac{2m_{\delta}}{r}\right)^{-1}dr^{2}+r^{2}\,d\Omega_{2}^{2}\label{eq: 5}
\end{equation}
Note that the line element (\ref{eq: 5}) can also be obtained as
the solution of the Einstein field equations with equation (\ref{eq: 2})
being the matter content. Thus, the event horizon can be obtained
by solving $g^{rr}(r_{h})=0$ as 
\begin{equation}
r_{h}=2m_{\delta}(r_{h})=\frac{4M}{\sqrt{\pi}}\,\gamma\left(\frac{3}{2},\frac{r_{h}^{2}}{4\delta}\right)\label{eq: 6}
\end{equation}
which clearly shows that $r_{h}$ cannot be solved in closed form.

For the sake of completeness, we stress that non-commutative corrections
to different thermodynamical quantities have been discussed in \cite{key-44}.

\section{Tunnelling approach to radiation spectrum: GUP corrections}

In strictly thermal approach, the probability of emission of Hawking
quanta is \cite{key-1} 
\begin{equation}
\Gamma\sim\mbox{exp}\left(-\frac{\omega}{T_{H}}\right)\label{eq: 7}
\end{equation}
where $\omega$ is the energy-frequency of the emitted radiation and
$T_{H}=\frac{1}{8\pi M}$ is the usual Hawking temperature. This probability
of emission has been modified by Parikh and Wilezek in the tunnelling
framework, considering contributions beyond the semi-classical approximation
as \cite{key-2,key-3} 
\begin{equation}
\Gamma\sim\mbox{exp}\left[-\frac{\omega}{T_{H}}\left(1-\frac{\omega}{2M}\right)\right]=\alpha\mbox{exp}\left[-\frac{\omega}{T_{H}}\left(1-\frac{\omega}{2M}\right)\right],\label{eq: 8}
\end{equation}
with $\alpha\sim1.$ For the sake of completeness, we stress that
in \cite{key-41} non thermal corrections to the Hawking effect were
discussed through a different approach. In that case, the structure
of $\Gamma$ is retained as $\mbox{exp}\left[-\beta_{(corr)}\omega\right]$,
where $\beta_{(corr)}$ is given by \cite{key-41} 
\begin{equation}
\beta_{(corr)}\equiv\beta_{H}\left(1+\sum_{i}\frac{\beta_{i}}{M^{2i}}\right).\label{eq: beta corretto}
\end{equation}
The non leading terms are the corrections to the temperature due to
quantum effects and $\beta_{H}\equiv T_{H}^{-1},$ see \cite{key-41}
for details. 

Now, comparing the above two probabilities in eqs. ($\ref{eq: 7}$)
and ($\ref{eq: 8})$ one can introduce the notion of the effective
BH temperature $T_{E}$ as \cite{key-14,key-15,key-19,key-20} 
\begin{equation}
T_{E}\equiv\frac{2M}{2M-1}\,T_{H}=\frac{1}{4\pi(2M-\omega)}.\label{eq: 9}
\end{equation}
Analogously, one can define the effective mass and effective horizon
radius as \cite{key-14,key-15,key-19,key-20} 
\begin{equation}
M_{E}=M-\frac{\omega}{2}\quad\mbox{and}\quad r_{E}=2M_{E}=2M-\omega\label{eq: 10}
\end{equation}
Further, these effective quantities can be interpreted as average
values of the corresponding quantities before ( \emph{i.e.} initial)
and after ( \emph{i.e.} final) the particle emission \cite{key-14,key-15,key-19,key-20}.
Accordingly, the effective temperature $T_{E}$ is the inverse of
the average value of the inverses of the initial and final Hawking
temperatures \cite{key-14,key-15,key-19,key-20}. As a result, the
effective Schwarzschild line element is \cite{key-15} 
\begin{equation}
ds^{2}=-\left(1-\frac{2M_{E}}{r}\right)dt^{2}+\left(1-\frac{2M_{E}}{r}\right)^{-1}dr^{2}+r^{2}\,d\Omega_{2}^{2}\label{eq: 11}
\end{equation}
One can describe this line element as the BH dynamical geometry during
the emission of the particle. Further, proceeding along the line of
approach of one of us (C. Corda) \cite{key-15}, the final non-strictly
thermal distributions considering the BH dynamical geometry take the
form \cite{key-15} 
\begin{eqnarray}
\left.\begin{array}{l}
\left\langle n\right\rangle _{b}=\frac{1}{\mbox{exp}\left[4\pi(2M-\omega)\omega\right]-1}\\
\left\langle n\right\rangle _{f}=\frac{1}{\mbox{exp}\left[4\pi(2M-\omega)\omega\right]+1}
\end{array}\right.\label{eq: 12}
\end{eqnarray}
where the suffices $b$ and $f$ represent boson and fermion particles
respectively. This tunnelling approach has been recently finalized
in \cite{key-27}. 

On the other hand, by taking into account the non-commutative geometry
discussed in Section 1 of this paper, starting from the non-commutative
form of the Schwarzschild metric (\ref{eq: 5}), eqs. from (\ref{eq: 7})
to (\ref{eq: 12}) must be replaced by their non-commutative counterparts
as 
\begin{equation}
\Gamma\sim\mbox{exp}\left(-\frac{\omega}{T_{H\delta}}\right),\label{eq: 13}
\end{equation}
where $T_{H\delta}=\frac{1}{8\pi m_{\delta}}$ is the Hawking temperature
in the non-commutative geometry, 
\begin{equation}
\Gamma\sim\mbox{exp}\left[-\frac{\omega}{T_{H\delta}}\left(1-\frac{\omega}{2m_{\delta}}\right)\right],\label{eq: 14}
\end{equation}
\begin{equation}
T_{E\delta}\equiv\frac{2m_{\delta}}{2m_{\delta}-1}\,T_{H\delta}=\frac{1}{4\pi(2m_{\delta}-\omega)},\label{eq: 15}
\end{equation}
\begin{equation}
m_{E\delta}=m_{\delta}-\frac{\omega}{2}\quad\mbox{and}\quad r_{E}=2m_{E\delta}=2m_{\delta}-\omega,\label{eq: 16}
\end{equation}
\begin{equation}
ds^{2}=-\left(1-\frac{2m_{E\delta}}{r}\right)dt^{2}+\left(1-\frac{2m_{E\delta}}{r}\right)^{-1}dr^{2}+r^{2}\,d\Omega_{2}^{2},\label{eq: 17}
\end{equation}
\begin{eqnarray}
\left.\begin{array}{l}
\left\langle n\right\rangle _{b}=\frac{1}{\mbox{exp}\left[4\pi(2m_{\delta}-\omega)\omega\right]-1}\\
\left\langle n\right\rangle _{f}=\frac{1}{\mbox{exp}\left[4\pi(2m_{\delta}-\omega)\omega\right]+1}
\end{array}\right..\label{eq: 18}
\end{eqnarray}
 Now, let us consider the modified Hawking temperature due to GUP
corrections \cite{key-21,key-22}. It can be expressed as \cite{key-21,key-22}
\begin{equation}
T_{H}^{(GUP)}=\frac{1}{8\pi M}\left[1-\frac{\alpha}{8\pi M}+5\left(\frac{\alpha}{8\pi M}\right)^{2}\right].\label{eq: Hawking temperature GUP}
\end{equation}
As a consequence, we can introduce \emph{a GUP modified BH mass and
a GUP modified horizon} \emph{radius} as \cite{key-21,key-22} 
\begin{equation}
\begin{array}{ccc}
M^{(GUP)}\equiv\frac{M}{\left[1-\frac{\alpha}{8\pi M}+5\left(\frac{\alpha}{8\pi M}\right)^{2}\right]} & \;and\; & r^{(GUP)}\equiv2M^{(GUP)},\end{array}\label{eq: GUP mass}
\end{equation}
respectively. Thus, eq. (\ref{eq: Hawking temperature GUP}) reads
\begin{equation}
T_{H}^{(GUP)}=\frac{1}{8\pi M^{(GUP)}}.\label{eq: Hawking temperature GUP 2}
\end{equation}
Using the Hawking's periodicity argument \cite{key-15,key-21,key-25,key-26,key-41}
one obtains the modified GUP Schwarzschild like line element \cite{key-21}
\begin{equation}
\left[ds^{(GUP)}\right]^{2}=-(1-\frac{2M^{(GUP)}}{r})dt^{2}+\frac{dr^{2}}{1-\frac{2M^{(GUP)}}{r}}+r^{2}(\sin^{2}\theta d\varphi^{2}+d\theta^{2})\label{eq: GUP Schwarzschild}
\end{equation}
and \cite{key-21}
\begin{equation}
\kappa{}^{(GUP)}\equiv\frac{1}{4M^{(GUP)}},\label{eq: modified surface gravity}
\end{equation}
as the \emph{GUP modified surface gravity}. The modified Schwarzschild
solution (\ref{eq: GUP Schwarzschild}) is obtained using the GUP
in the background of non-commutative geometry. Thus, it is clear that
the GUP Schwarzschild solution is not a solution of the Einstein field
equation. It will be interesting to find modification of Einstein
gravity for which the GUP Schwarzschild solution is a solution. This
could be the subject of future works. 

Now, combining this GUP correction with the notion of effective temperature,
one can introduce the \emph{GUP corrected effective temperature }as\emph{
}\cite{key-21}
\begin{equation}
T_{E}^{(GUP)}(\omega)\equiv\frac{2M^{(GUP)}}{2M^{(GUP)}-\omega}T_{H}^{(GUP)}=\frac{1}{4\pi(2M^{(GUP)}-\omega)},
\end{equation}
the \emph{GUP corrected effective Boltzmann factor }as \cite{key-21}\emph{
}
\begin{equation}
\beta_{E}^{(GUP)}(\omega)\equiv\frac{1}{T_{E}^{(GUP)}(\omega)}
\end{equation}
and the \emph{GUP corrected effective mass} and \emph{effective horizon}
\emph{radius }as \cite{key-21} 
\begin{equation}
M_{E}^{(GUP)}=M^{(GUP)}-\frac{\omega}{2}\qquad and\qquad r_{E}^{(GUP)}=2M_{E}^{(GUP)}=2M^{(GUP)}-\omega.\label{eq: GUP corrected effective mass}
\end{equation}
Note that the equations from (\ref{eq: Hawking temperature GUP})
to (\ref{eq: modified surface gravity}) represent GUP corrections
using Hamilton-Jacobi method beyond the semi-classical approximation.
By taking into account the non-commutative geometry, the equations
from (\ref{eq: Hawking temperature GUP}) to (\ref{eq: GUP corrected effective mass})
become 
\begin{equation}
T_{H\delta}^{(GUP)}=\frac{1}{8\pi m_{\delta}}\left[1-\frac{\alpha}{8\pi m_{\delta}}+5\left(\frac{\alpha}{8\pi m_{\delta}}\right)^{2}\right],\label{eq: Hawking temperature GUP NC}
\end{equation}
 
\begin{equation}
\begin{array}{ccc}
m_{\delta}^{(GUP)}\equiv\frac{m_{\delta}}{\left[1-\frac{\alpha}{8\pi m_{\delta}}+5\left(\frac{\alpha}{8\pi m_{\delta}}\right)^{2}\right]}, &  & r_{\delta}^{(GUP)}\equiv2m_{\delta}^{(GUP)}\end{array},\label{eq: GUP mass NC}
\end{equation}
 
\begin{equation}
T_{H\delta}^{(GUP)}=\frac{1}{8\pi m_{\delta}^{(GUP)}},\label{eq: Hawking temperature GUP NC 2}
\end{equation}
 
\begin{equation}
\left[ds_{\delta}^{(GUP)}\right]^{2}=-(1-\frac{2m_{\delta}^{(GUP)}}{r})dt^{2}+\frac{dr^{2}}{1-\frac{2m_{\delta}^{(GUP)}}{r}}+r^{2}(\sin^{2}\theta d\varphi^{2}+d\theta^{2}),\label{eq: GUP Schwarzschild NC}
\end{equation}
\begin{equation}
\kappa_{\delta}{}^{(GUP)}\equiv\frac{1}{4m_{\delta}^{(GUP)}},\label{eq: modified surface gravity GUP NC}
\end{equation}
\begin{equation}
T_{E\delta}^{(GUP)}(\omega)\equiv\frac{2m_{\delta}^{(GUP)}}{2m_{\delta}^{(GUP)}-\omega}T_{H_{\delta}}^{(GUP)}=\frac{1}{4\pi(2m_{\delta}^{(GUP)}-\omega)},
\end{equation}
\emph{ }
\begin{equation}
\beta_{E\delta}^{(GUP)}(\omega)\equiv\frac{1}{T_{E\delta}^{(GUP)}(\omega)}
\end{equation}
 
\begin{equation}
m_{E\delta}^{(GUP)}=m_{\delta}^{(GUP)}-\frac{\omega}{2}\qquad and\qquad r_{E\delta}^{(GUP)}=2m_{E\delta}^{(GUP)}=2m_{\delta}^{(GUP)}-\omega.\label{eq: GUP corrected effective mass NC}
\end{equation}
Further, by Hawking's periodicity argument \cite{key-15,key-21,key-25,key-26}
one easily obtains the modified GUP non-commutative effective Schwarzschild
like line element as 
\begin{equation}
\left[ds_{E\delta}^{(GUP)}\right]^{2}=-(1-\frac{2m_{E\delta}^{(GUP)}}{r})dt^{2}+\frac{dr^{2}}{1-\frac{2m_{E\delta}^{(GUP)}}{r}}+r^{2}(\sin^{2}\theta d\varphi^{2}+d\theta^{2}),\label{eq: effective GUP Schwarzschild NC}
\end{equation}
Now, if one follows step by step the analysis in \cite{key-15,key-21},
then at the end one obtains the correct physical states for boson
and fermions as 
\begin{equation}
\begin{array}{c}
|\Psi>_{boson}=\left(1-\exp\left(-8\pi m_{E\delta}^{(GUP)}\omega\right)\right)^{\frac{1}{2}}\sum_{n}\exp\left(-4\pi nm_{E\delta}^{(GUP)}\omega\right)|n_{out}^{(L)}>\otimes|n_{out}^{(R)}>\\
\\
|\Psi>_{fermion}=\left(1+\exp\left(-8\pi m_{E\delta}^{(GUP)}\omega\right)\right)^{-\frac{1}{2}}\sum_{n}\exp\left(-4\pi nm_{E\delta}^{(GUP)}\omega\right)|n_{out}^{(L)}>\otimes|n_{out}^{(R)}>
\end{array}\label{eq: physical states-1}
\end{equation}
and the correct distributions as 
\begin{equation}
\begin{array}{c}
<n>_{boson}=\frac{1}{\exp\left(8\pi m_{E\delta}^{(GUP)}\omega\right)-1}=\frac{1}{\exp\left[4\pi\left(2m_{\delta}^{(GUP)}-\omega\right)\omega\right]-1}\\
\\
<n>_{fermion}=\frac{1}{\exp\left(8\pi m_{E\delta}^{(GUP)}\omega\right)+1}=\frac{1}{\exp\left[4\pi\left(2m_{\delta}^{(GUP)}-\omega\right)\omega\right]+1},
\end{array}\label{eq: final distributions}
\end{equation}
which, by using the first of eqs. (\ref{eq: GUP mass NC}) become
\begin{equation}
\begin{array}{c}
<n>_{boson}=\frac{1}{\exp\left[4\pi\left(\frac{m_{\delta}}{\left[1-\frac{\alpha}{8\pi m_{\delta}}+5\left(\frac{\alpha}{8\pi m_{\delta}}\right)^{2}\right]}-\omega\right)\omega\right]-1}\\
\\
<n>_{fermion}=\frac{1}{\exp\left[4\pi\left(\frac{m_{\delta}}{\left[1-\frac{\alpha}{8\pi m_{\delta}}+5\left(\frac{\alpha}{8\pi m_{\delta}}\right)^{2}\right]}-\omega\right)\omega\right]+1},
\end{array}\label{eq: final distributions 2}
\end{equation}
The above expressions of the distributions clearly show that they
are not thermal in nature as both the BH dynamical geometry during
the emission of the particle, the GUP corrections to the semi classical
Hawking temperature and the non-commutative geometry are taken into
account. 

\section{Corrections to the Bohr-like black hole}

The general conviction that BHs should be highly excited states representing
the fundamental bricks of quantum gravity \cite{key-28} has been
shown to be correct in the recent works {[}13,18-22{]}. In such papers,
one of us (C. Corda) has indeed shown that the Schwarzschild BH is
the gravitational analogous of the historical semi-classical Bohr's
hydrogen atom \cite{key-33,key-34}. The Bohr-like approach to BH
quantum physics started with the pioneering works \cite{key-19,key-20}.
It works through the natural correspondence between Hawking radiation
and BH quasi-normal modes (QNMs) {[}13,18-22{]}. Considering an isolated
BH (in the same way that Bohr considered an isolated hydrogen atom),
the emissions of Hawking quanta and the absorptions of external particles
``trigger'' the BH QNMs {[}13,18-22{]}. In this analogy, BH QNMs
represent the \textquotedbl{}electron\textquotedbl{} jumping from
a quantum level to another one. Hence, their absolute values are the
energy \textquotedbl{}shells\textquotedbl{} of the ``gravitational
hydrogen atom'' {[}13,18-22{]}. Remarkably, the time evolution of
the system permits to solve the BH information puzzle \cite{key-14,key-30}.
The results in {[}13,18-22{]} are also consistent with previous results
in the literature, included the historic result of Bekenstein on the
area quantization \cite{key-35}. The Bohr-like framework for BH quantum
physics also finalizes the famous tunnelling approach of Parikh and
Wilczek. One indeed finds the correct value of the pre-factor of the
Parikh and Wilczek probability of emission, i.e. $\alpha$ in eq.
(\ref{eq: 8}), as \cite{key-27} 
\begin{equation}
\alpha\equiv\alpha_{m}=\frac{1-\exp\left[-2\pi\right]}{1-\exp\left[-2\pi\left(n_{max}-m+1\right)\right]}.\label{eq: mio alpha number}
\end{equation}
 In this equation $n_{max}$ represents the maximum value of the principal
quantum number $n$ which can be found in Eq. (17) in \cite{key-27},
while $m$ is the BH excited level. Thus, $\alpha$ depends on the
BH quantum level, see \cite{key-27} for details. This result permits
to write down the probability of emission between two generic BH quantum
levels $m$ and $n$ in the intriguing form \cite{key-27}

\begin{equation}
\begin{array}{c}
\Gamma_{m\rightarrow n}=\alpha_{m}\exp-\left\{ \frac{\Delta E_{m\rightarrow n}}{\left[T_{E}(\omega)\right]_{m\rightarrow n}}\right\} =\alpha_{m}\exp\left[-2\pi\left(n-m\right)\right]=\\
\\
=\left\{ \frac{1-\exp\left[-2\pi\right]}{1-\exp\left[-2\pi\left(n_{max}-m+1\right)\right]}\right\} \exp\left[-2\pi\left(n-m\right)\right].
\end{array}\label{eq: Corda Probability Intriguing finalized}
\end{equation}
In a quantum mechanical framework, Hawking radiation can be physically
interpreted in terms of quantum jumps among unperturbed levels {[}13,16-22{]}. 

For large values of $n,$ that is for excited BHs, the QNMs' expression
of the Schwarzschild BH is independent of the angular momentum quantum
number {[}13,16-22{]}. In order to take into account the non-strictly
thermality of the radiation spectrum, one replaces the Hawking temperature
with the effective temperature in the standard (thermal) QNMs equation
obtaining {[}13,16-22{]}
\begin{equation}
\begin{array}{c}
\omega_{n}=a+ib+2\pi in\times T_{E}(|\omega_{n}|)=\\
\\
\backsimeq2\pi in\times T_{E}(|\omega_{n}|)=\frac{in}{4M-2|\omega_{n}|}.
\end{array}\label{eq: quasinormal modes corrected}
\end{equation}
Here $a$ and $b$ are real numbers with $a=(\ln3)\times T_{E}(|\omega_{n}|),\;b=\pi\times T_{E}(|\omega_{n}|)$
for $j=0,2$ (scalar and gravitational perturbations), $a=0,\;b=0$
for $j=1$ (vector perturbations) and $a=0,\;b=\pi\times T_{E}(|\omega_{n}|)$
for half-integer values of $j$. 

Now, let us see how the corrections due to the GUP and to non-commutative
geometry change the model. As we use the modified GUP in non-commutative
effective Schwarzschild line element (\ref{eq: effective GUP Schwarzschild NC})
instead of the effective Schwarzschild line element (\ref{eq: 11}),
eq. (\ref{eq: quasinormal modes corrected}) must be replaced by 
\begin{equation}
\begin{array}{c}
\omega_{n}=a+ib+2\pi in\times T_{E\delta}^{(GUP)}(|\omega_{n}|)=\\
\\
\backsimeq2\pi in\times T_{E\delta}^{(GUP)}(|\omega_{n}|)=\frac{in}{4m_{\delta}^{(GUP)}-2|\omega_{n}|}=\\
\\
=\frac{in}{4\frac{m_{\delta}}{\left[1-\frac{\alpha}{8\pi m_{\delta}}+5\left(\frac{\alpha}{8\pi m_{\delta}}\right)^{2}\right]}-2|\omega_{n}|}.
\end{array}\label{eq: quasinormal modes corrected GUP NC}
\end{equation}
The solution of (\ref{eq: quasinormal modes corrected GUP NC}) in
terms of $|\omega_{n}|$ reads 

\begin{equation}
\begin{array}{c}
|\omega_{n}|=\frac{m_{\delta}}{\left[1-\frac{\alpha}{8\pi m_{\delta}}+5\left(\frac{\alpha}{8\pi m_{\delta}}\right)^{2}\right]}+\\
\\
\pm\sqrt{\left(\frac{m_{\delta}}{\left[1-\frac{\alpha}{8\pi m_{\delta}}+5\left(\frac{\alpha}{8\pi m_{\delta}}\right)^{2}\right]}\right)^{2}-\frac{n}{2}}.
\end{array}\label{eq: doppia soluzione}
\end{equation}
Clearly, a BH does not emit more energy than its total mass {[}13,16-22{]}.
In this case, we must take into account the corrections to the mass
arising from the GUP and non-commutative geometry. Thus, the physical
solution is the one obeying 
\begin{equation}
|\omega_{n}|<\frac{m_{\delta}}{\left[1-\frac{\alpha}{8\pi m_{\delta}}+5\left(\frac{\alpha}{8\pi m_{\delta}}\right)^{2}\right]},\label{eq: obbedienza}
\end{equation}
that is 
\begin{equation}
\begin{array}{c}
E_{n}\equiv|\omega_{n}|=\frac{m_{\delta}}{\left[1-\frac{\alpha}{8\pi m_{\delta}}+5\left(\frac{\alpha}{8\pi m_{\delta}}\right)^{2}\right]}+\\
\\
-\sqrt{\left(\frac{m_{\delta}}{\left[1-\frac{\alpha}{8\pi m_{\delta}}+5\left(\frac{\alpha}{8\pi m_{\delta}}\right)^{2}\right]}\right)^{2}-\frac{n}{2}}.
\end{array}\label{eq: radice fisica}
\end{equation}
$E_{n}\:$ is the value of total energy emitted by the BH when it
is excited at a level $n$ {[}13,16-22{]}.

\noindent \quad{}\quad{}Let us consider an emission from the ground
state (i.e. a BH which is not excited) to a state with large $n=n_{1}$.
Then, by using Eq. (\ref{eq: radice fisica}), the GUP corrected mass
of the analogous BH changes from $m_{\delta}^{(GUP)}\:$ to 

\begin{equation}
\begin{array}{c}
m_{\delta}^{(GUP)}{}_{n_{1}}\equiv m_{\delta}^{(GUP)}-E_{n_{1}}=\\
\\
=\sqrt{m_{\delta}^{(GUP)}{}^{2}-\frac{n_{1}}{2}}=\sqrt{\left(\frac{m_{\delta}}{\left[1-\frac{\alpha}{8\pi m_{\delta}}+5\left(\frac{\alpha}{8\pi m_{\delta}}\right)^{2}\right]}\right){}^{2}-\frac{n_{1}}{2}}.
\end{array}\label{eq: me-1}
\end{equation}
If now one considers a transition from $n=n_{1}$ to a different state
with $n=n_{2}$, having $n_{2}>n_{1}$, the GUP corrected mass of
the analogous BH changes again from $m_{\delta}^{(GUP)}{}_{n_{1}}\:$
to 

\begin{equation}
\begin{array}{c}
m_{\delta}^{(GUP)}{}_{n_{2}}\equiv m_{\delta}^{(GUP)}{}_{n_{1}}-\Delta E_{n_{1}\rightarrow n_{2}}=m_{\delta}^{(GUP)}-E_{n_{2}}=\\
=\sqrt{\left(\frac{m_{\delta}}{\left[1-\frac{\alpha}{8\pi m_{\delta}}+5\left(\frac{\alpha}{8\pi m_{\delta}}\right)^{2}\right]}\right){}^{2}-\frac{n_{2}}{2}}.
\end{array}\label{eq: me}
\end{equation}
$\Delta E_{n_{1}\rightarrow n_{2}}$ in previous equation is given
by
\begin{equation}
\begin{array}{c}
\Delta E_{n_{1}\rightarrow n_{2}}\equiv E_{n_{2}}-E_{n_{1}}=m_{\delta}^{(GUP)}{}_{n_{1}}-m_{\delta}^{(GUP)}{}_{n_{2}}=\\
\\
=\sqrt{\left(\frac{m_{\delta}}{\left[1-\frac{\alpha}{8\pi m_{\delta}}+5\left(\frac{\alpha}{8\pi m_{\delta}}\right)^{2}\right]}\right){}^{2}-\frac{n_{1}}{2}}-\sqrt{\left(\frac{m_{\delta}}{\left[1-\frac{\alpha}{8\pi m_{\delta}}+5\left(\frac{\alpha}{8\pi m_{\delta}}\right)^{2}\right]}\right){}^{2}-\frac{n_{2}}{2}}
\end{array},\label{eq: jump}
\end{equation}
and represents the jump between the two levels due to the emission
of a particle having frequency $\Delta E_{n_{1}\rightarrow n_{2}}$.
Such a discrete amount of energy corresponds to a quantum jump. The
issue that for large $n$ one finds independence on the other quantum
numbers is perfectly consistent with \emph{the Correspondence Principle}
stated by Bohr \cite{key-36}. This principle indeed claims that \textquotedblleft transition
frequencies at large quantum numbers should equal classical oscillation
frequencies\textquotedblright . We stress again the analogy with Bohr's
hydrogen atom. In fact, in that model \cite{key-33,key-34} energy
is gained and loss by electrons through quantum jumps from one allowed
energy shell to another. Hence, radiation can be absorbed or emitted
and the energy difference of the levels respects the Planck relation
$E=hf$ (in standard units), where $\:h\:$ is the Planck constant
and $f\:$ is the transition frequency. In the present GUP corrected
analogous Bohr-like BH, QNMs (the ``gravitational electrons'') only
gain and lose energy through jumps from one allowed energy shell to
another with absorptions or emissions of Hawking quanta, but now the
energy difference of the levels is governed by Eq. (\ref{eq: jump}).
Remarkably, one interpretes Eq. (\ref{eq: radice fisica}) in terms
of a particle, the ``electron'', which is quantized on a circle
of length 
\begin{equation}
\begin{array}{c}
L=\frac{1}{T_{E\delta}^{(GUP)}(E_{n})}=\\
\\
=4\pi\left[\frac{m_{\delta}}{\left[1-\frac{\alpha}{8\pi m_{\delta}}+5\left(\frac{\alpha}{8\pi m_{\delta}}\right)^{2}\right]}+\sqrt{\left(\frac{m_{\delta}}{\left[1-\frac{\alpha}{8\pi m_{\delta}}+5\left(\frac{\alpha}{8\pi m_{\delta}}\right)^{2}\right]}\right)^{2}-\frac{n}{2}}\right].
\end{array}\label{eq: lunghezza cerchio}
\end{equation}
This finalizes the cited similarity with Bohr's hydrogen atom. Eq.
(\ref{eq: lunghezza cerchio}) represents indeed a perfect analogy
with the electron travelling around the hydrogen nucleus with circular
orbits in Bohr's approach \cite{key-33,key-34} and is also similar
to planets travelling around the Sun in our solar system. We stress
that Bohr's hydrogen atom represents an approximated model with respect
to the valence shell atom model of full quantum mechanics. In the
same way, the present GUP corrected analogous Bohr-like BH should
be a better approximated model with respect to previous results in
{[}13,18-21{]}. But it is still far from the final, currently unknown,
BH model of a full unitary quantum gravity theory. 

Now, let us set $n_{1}=n-1$, $n_{2}=n$ in eq. (\ref{eq: jump}).
We get the emitted energy for a jump between two neighboring levels
as 
\begin{equation}
\begin{array}{c}
\Delta E_{n-1\rightarrow n}=\sqrt{\left(\frac{m_{\delta}}{\left[1-\frac{\alpha}{8\pi m_{\delta}}+5\left(\frac{\alpha}{8\pi m_{\delta}}\right)^{2}\right]}\right)^{2}-\frac{n-1}{2}}+\\
\\
-\sqrt{\left(\frac{m_{\delta}}{\left[1-\frac{\alpha}{8\pi m_{\delta}}+5\left(\frac{\alpha}{8\pi m_{\delta}}\right)^{2}\right]}\right)^{2}-\frac{n}{2}}.
\end{array}\label{eq: variazione}
\end{equation}
Bekenstein \cite{key-35} has shown that the Schwarzschild BH area
quantum should be $\triangle A=8\pi$ (the \emph{Planck length} $l_{p}=1.616\times10^{-33}\mbox{ }cm$
is equal to one in Planck units). In Schwarzschild BHs the \emph{horizon
area} $A$ is connected to the mass by the relation $A=16\pi M^{2}.$
Hence, a variation $\triangle M\,$ of the mass enables the variation

\noindent 
\begin{equation}
\triangle A=32\pi M\triangle M\label{eq: variazione area}
\end{equation}
of the area. Setting $\triangle M=-\Delta E_{n-1\rightarrow n}$ (the
case of an emission) if one uses Eqs. (\ref{eq: me-1}) and (\ref{eq: variazione})
one gets 
\begin{equation}
\triangle A_{n-1}\equiv-32\pi m_{\delta}^{(GUP)}{}_{n-1}\Delta E_{n-1\rightarrow n}.\label{eq: area quantum e}
\end{equation}
One can think that Eq. (\ref{eq: area quantum e}) gives the area
quantum of an excited GUP corrected analogous BH for a jump from the
level $n-1$ to the level $n$ in function of the ``overtone'' number
$n$ and of the initial GUP corrected analogous BH mass. But we see
that one has a problem using eq. (\ref{eq: area quantum e}). An absorption
from the level $n$ to the level $n-1$ is indeed possible, through
the absorbed energy 
\begin{equation}
\begin{array}{c}
\Delta E_{n\rightarrow n-1}=-\Delta E_{n-1\rightarrow n}=\\
\\
=\sqrt{\left(\frac{m_{\delta}}{\left[1-\frac{\alpha}{8\pi m_{\delta}}+5\left(\frac{\alpha}{8\pi m_{\delta}}\right)^{2}\right]}\right)^{2}-\frac{n}{2}}-\sqrt{\left(\frac{m_{\delta}}{\left[1-\frac{\alpha}{8\pi m_{\delta}}+5\left(\frac{\alpha}{8\pi m_{\delta}}\right)^{2}\right]}\right)^{2}-\frac{n-1}{2}}.
\end{array}\label{eq: absorbed}
\end{equation}
Hence, if one sets $\triangle M=-\Delta E_{n\rightarrow n-1}=\Delta E_{n-1\rightarrow n}$
one gets a quantum of area 
\begin{equation}
\triangle A_{n}\equiv-32\pi m_{\delta}^{(GUP)}{}_{n}\Delta E_{n\rightarrow n-1}=32\pi m_{\delta}^{(GUP)}{}_{n}\Delta E_{n-1\rightarrow n}.\label{eq: area quantum a}
\end{equation}
Thus, one finds that the absolute value of the area quantum for an
absorption between two levels is not the same as the absolute value
of the area quantum for an emission between the same levels. This
is because $m_{\delta}^{(GUP)}{}_{n}\neq m_{\delta}^{(GUP)}{}_{n-1}$.
Instead, we intuitively expect the area spectrum to be the equal for
absorption and emission. \cite{key-14,key-30}. One solves this problem
considering the \emph{GUP corrected analogous effective mass} which
corresponds to the transitions between the two levels $n\;$ and $n-1$.
This latter is indeed the same for emission and absorption: 
\begin{equation}
\begin{array}{c}
m_{E\delta(n,\;n-1)}^{(GUP)}\equiv\frac{1}{2}\left(\frac{m_{\delta}}{\left[1-\frac{\alpha}{8\pi m_{\delta}}+5\left(\frac{\alpha}{8\pi m_{\delta}}\right)^{2}\right]}_{n-1}+\frac{m_{\delta}}{\left[1-\frac{\alpha}{8\pi m_{\delta}}+5\left(\frac{\alpha}{8\pi m_{\delta}}\right)^{2}\right]}_{n}\right)=\\
\\
=\frac{1}{2}\left(\sqrt{\left(\frac{m_{\delta}}{\left[1-\frac{\alpha}{8\pi m_{\delta}}+5\left(\frac{\alpha}{8\pi m_{\delta}}\right)^{2}\right]}\right)^{2}-\frac{n-1}{2}}+\sqrt{\left(\frac{m_{\delta}}{\left[1-\frac{\alpha}{8\pi m_{\delta}}+5\left(\frac{\alpha}{8\pi m_{\delta}}\right)^{2}\right]}\right)^{2}-\frac{n}{2}}\right).
\end{array}\label{eq: massa effettiva n}
\end{equation}
Hence, let us replace $m_{\delta}^{(GUP)}{}_{n-1}$ with $m_{E\delta(n,\;n-1)}^{(GUP)}$
in eq. (\ref{eq: area quantum e}) and $m_{\delta}^{(GUP)}{}_{n}$
again with $M_{E(n,\;n-1)}$ in eq. (\ref{eq: area quantum a}). We
find 
\begin{equation}
\begin{array}{c}
\triangle A_{n-1}\equiv-32\pi m_{E\delta(n,\;n-1)}^{(GUP)}\Delta E_{n-1\rightarrow n}\qquad emission\\
\\
\triangle A_{n}\equiv-32\pi m_{E\delta(n,\;n-1)}^{(GUP)}\Delta E_{n\rightarrow n-1}\qquad absorption.
\end{array}\label{eq: expects}
\end{equation}
Thus, now one gets $|\triangle A_{n}|=|\triangle A_{n-1}|.$ Eqs.
(\ref{eq: absorbed}), (\ref{eq: massa effettiva n}) and some algebra
give

\begin{equation}
|\triangle A_{n}|=|\triangle A_{n-1}|=8\pi.\label{eq: 8 pi planck}
\end{equation}
Thus, we find the very intriguing result that the famous law of Bekenstein
on the area quantization \cite{key-35} is affected neither by non-commutative
geometry nor by the GUP. This is a clear indication of the universality
of Bekenstein's result on the area quantization \cite{key-35} . 

If one puts $A_{n-1}\equiv16\pi\left(m_{\delta}^{(GUP)}{}_{n-1}\right)^{2}$,
$A_{n}\equiv16\pi\left(m_{\delta}^{(GUP)}{}_{n}\right)^{2},$ one
finds the formulas of the number of quanta of area as 
\begin{equation}
\begin{array}{c}
N_{n-1}\equiv\frac{A_{n-1}}{|\triangle A_{n-1}|}=\\
\\
=\frac{16\pi\left(m_{\delta}^{(GUP)}{}_{n-1}\right)^{2}}{32\pi m_{E\delta(n,\;n-1)}^{(GUP)}\cdot\Delta E_{n-1\rightarrow n}}=\frac{\left(m_{\delta}^{(GUP)}{}_{n-1}\right)^{2}}{2m_{E\delta(n,\;n-1)}^{(GUP)}\cdot\Delta E_{n-1\rightarrow n}}
\end{array}\label{eq: N n-1}
\end{equation}

\noindent before the emission, and
\begin{equation}
\begin{array}{c}
N_{n}\equiv\frac{A_{n}}{|\triangle A_{n}|}=\\
\\
=\frac{16\pi\left(m_{\delta}^{(GUP)}{}_{n}\right)^{2},}{32\pi m_{E\delta(n,\;n-1)}^{(GUP)}\cdot\Delta E_{n-1\rightarrow n}}=\frac{\left(m_{\delta}^{(GUP)}{}_{n}\right)^{2}}{2m_{E\delta(n,\;n-1)}^{(GUP)}\cdot\Delta E_{n-1\rightarrow n}}
\end{array}\label{eq: N n}
\end{equation}
after the emission respectively. This implies 

\noindent 
\begin{equation}
\begin{array}{c}
N_{n}-N_{n-1}=\frac{\left(m_{\delta}^{(GUP)}{}_{n}\right)^{2}-\left(m_{\delta}^{(GUP)}{}_{n-1}\right)^{2},}{2m_{E\delta(n,\;n-1)}^{(GUP)}\cdot\Delta E_{n-1\rightarrow n}}=\\
\\
=\frac{\Delta E_{n-1\rightarrow n}\left(m_{\delta}^{(GUP)}{}_{n-1}+m_{\delta}^{(GUP)}{}_{n}\right)}{2m_{E\delta(n,\;n-1)}^{(GUP)}\cdot\Delta E_{n-1\rightarrow n}}=1,
\end{array}\label{eq: check}
\end{equation}
as one expects. Now, one can write down the famous formula of Bekenstein-Hawking
entropy \cite{key-1,key-37,key-38} as
\begin{equation}
\begin{array}{c}
\left(S_{BH}\right)_{n-1}\equiv\frac{A_{n-1}}{4}=8\pi N_{n-1}m_{\delta}^{(GUP)}{}_{n-1}\cdot\Delta E_{n-1\rightarrow n}=\\
\\
=4\pi\left[\left(m_{\delta}^{(GUP)}\right)^{2}-\frac{n+1}{2}\right]
\end{array}\label{eq: Bekenstein-Hawking  n-1}
\end{equation}
before the emission and 
\begin{equation}
\begin{array}{c}
\left(S_{BH}\right)_{n}\equiv\frac{A_{n}}{4}=8\pi N_{n}m_{\delta}^{(GUP)}{}_{n-1}\cdot\Delta E_{n-1\rightarrow n}=\\
\\
=4\pi\left[\left(m_{\delta}^{(GUP)}\right)^{2}-\frac{n}{2}\right]
\end{array}\label{eq: Bekenstein-Hawking  n}
\end{equation}
after the emission respectively. Hence, the Bekenstein-Hawking entropy
can be written as a function of the QNMs principal quantum number,
i.e. of the BH quantum excited state. Eqs. (\ref{eq: Bekenstein-Hawking  n-1})
and (\ref{eq: Bekenstein-Hawking  n}) permits to generalize the results
in {[}13,18-22{]} to the current case of the GUP corrected mass of
the analogous BH.

\noindent \quad{}\quad{}Now, we recall that the Bekenstein-Hawking
entropy cannot be considered the definitive answer for a correct quantum
theory of gravity \cite{key-39}. It is indeed very important going
beyond the Bekenstein-Hawking entropy and finding its sub-leading
corrections \cite{key-39}. By using the quantum tunnelling approach
one remarkably arrives to the sub-leading corrections at third order
approximation \cite{key-40} 
\begin{equation}
S_{total}=S_{BH}-\ln S_{BH}+\frac{3}{2A}+\frac{2}{A^{2}}.\label{eq: entropia totale}
\end{equation}
In this approach the total BH entropy depends on four different parts:
the standard Bekenstein-Hawking entropy, a logarithmic term, an inverse
area term and an inverse squared area term \cite{key-40}. Thus, one
can find the formulas of the total BH entropy taking into account
the sub-leading corrections at third order approximation and considering
the GUP corrected mass of the analogous BH as

\noindent 
\begin{equation}
\begin{array}{c}
\left(S_{total}\right)_{n-1}=4\pi\left[\left(m_{\delta}^{(GUP)}\right)^{2}-\frac{n-1}{2}\right]-\ln4\pi\left[\left(m_{\delta}^{(GUP)}\right)^{2}-\frac{n-1}{2}\right]+\\
\\
+\frac{3}{32\pi\left[\left(m_{\delta}^{(GUP)}\right)^{2}-\frac{n-1}{2}\right]}+\frac{2}{16\pi\left[\left(m_{\delta}^{(GUP)}\right)^{2}-\frac{n-1}{2}\right]^{2}}
\end{array}\label{eq: entropia n-1}
\end{equation}

\noindent before the emission, and 
\begin{equation}
\begin{array}{c}
\left(S_{total}\right)_{n}=4\pi\left[\left(m_{\delta}^{(GUP)}\right)^{2}-\frac{n}{2}\right]-\ln4\pi\left[\left(m_{\delta}^{(GUP)}\right)^{2}-\frac{n}{2}\right]+\\
\\
+\frac{3}{32\pi\left[\left(m_{\delta}^{(GUP)}\right)^{2}-\frac{n}{2}\right]}+\frac{2}{16\pi\left[\left(m_{\delta}^{(GUP)}\right)^{2}-\frac{n}{2}\right]^{2}}
\end{array}\label{eq: entropia n}
\end{equation}

\noindent after the emission, respectively. Therefore, the total BH
entropy at third order approximation can be written as a function
of the BH excited state $n.$ Again, here we improve the results in
{[}13,18-22{]} to the current case of the GUP corrected mass of the
analogous BH.

\section{Summary and Concluding Remarks}

The present work considers GUP correction of non-thermal radiation
spectrum in the background of non-commutative geometry using the framework
of tunnelling mechanism. At first, we formulated the line element
for Schwarzschild BH in the context of non-commutative geometry. Then
we introduced the notion of effective temperature, effective mass
and effective horizon radius considering contributions beyond semi
classical approximation. Also, following the idea of one of us, C.
Corda, we have determined the non-strictly thermal distributions for
bosons and fermions. Subsequently, we introduce the GUP correction
to the BH dynamical geometry by finding the final distributions for
the GUP corrected mass of the analogous BH. 

After that, it has been shown that the GUP and the non-commutative
geometry modify the Bohr-like BH model, recently discussed in {[}13,18-22{]}.
In particular, we found the intriguing result that the famous law
of Bekenstein on the area quantization \cite{key-35} is affected
neither by non-commutative geometry nor by the GUP. This is a clear
indication of the universality of Bekenstein's result. Finally, it
has been shown that both the Bekentsein-Hawking entropy and the total
BH entropy to third order approximation are still functions of the
BH quantum level, generalizing the results in {[}13,18-22{]}.

\section{Acknowledgements}

The author SC acknowledges IUCAA, Pune, India for their warm hospitality
and research facilities at Library. Also SC acknowledges the UGC-DRS
Programme in the Department of Mathematics, Jadavpur University. The
author SH is thankful to UGC NET-JRF for awarding Research fellowship.
The author CC has been supported financially by the Research Institute
for Astronomy and Astrophysics of Maragha (RIAAM), Project Number
1/4717-112. 

The authors thank the Editors and the unknown referees for useful
comments.


\begin{thebibliography}{10}
\bibitem{key-1} S. W. Hawking, Commun. Math. Phys. \textbf{43}, 199
(1975).

\bibitem{key-2}M. K. Parikh, F. Wilczek, Phys. Rev. Lett. \textbf{85},
5042 (2000). 

\bibitem{key-3}M. K. Parikh, Gen. Rel. Grav. \textbf{36}, 2419 (2004). 

\bibitem{key-4}R. Banerjee, B. R. Majhi, J. High Energ. Phys. \textbf{0806},
095 (2008). 

\bibitem{key-5}M. Angheben, M. Nadalini, L. Vanzo, S. Zerbini J.
High Energ. Phys. \textbf{0505}, 014 (2008). 

\bibitem{key-6}M. Arzano, A. J. M. Medved, E. C. Vagenas, J. High
Energ. Phys. \textbf{0509}, 037 (2005). 

\bibitem{key-7}R. Banerjee, B. R. Majhi, Phys. Lett. B \textbf{675},
243 (2009). 

\bibitem{key-8}Q. Q. Jiang, S. Q. Wu, X. Cai, Phys. Rev. D \textbf{73},
064003 (2006). 

\bibitem{key-9}Q. Q. Jiang, S. Q. Wu, X. Cai, Phys. Rev. D \textbf{73},
069902 (2006). 

\bibitem{key-10}R. Kerner, R. B. Mann, Phys. Rev. D 73, 104010 (2006). 

\bibitem{key-11}L. Vanzo, G. Acquaviva, R. di Criscienzo, Class.
Quant. Grav. \textbf{28}, 183001 (2011). 

\bibitem[12]{key-13}B. Zhang, Q. Y. Cai, M. S. Zhan, L. You, Int.
J. Mod. Phys. D 22, 1341014 (2013). 

\bibitem[13]{key-14}C. Corda, Ann. Phys. \textbf{353}, 71 (2015). 

\bibitem[14]{key-12}S. W. Hawking, Phys. Rev. D \textbf{14}, 2460
(1976).

\bibitem[15]{key-15}C. Corda, Ann. Phys. \textbf{337}, 49 (2013). 

\bibitem[16]{key-19}C. Corda, J. High Energ. Phys. \textbf{101},
1108 (2011).

\bibitem[17]{key-20}C. Corda, Int. J. Mod. Phys. D \textbf{21}, 1242023
(2012). 

\bibitem[18]{key-27}C. Corda, Class. Quantum Grav. 32, 195007 (2015).

\bibitem[19]{key-29}C. Corda, Eur. Phys. J. C 73, 2665 (2013). 

\bibitem[20]{key-30}C. Corda, Adv. High En. Phys. 867601 (2015). 

\bibitem[21]{key-31}C. Corda, Int. Journ. Theor. Phys. 54, 3841 (2015).

\bibitem[22]{key-32}C. Corda, AIP Conf. Proc. 1648, 020004 (2015).

\bibitem[23]{key-35}J. D. Bekenstein, Lett. Nuovo Cim. 11, 467 (1974). 

\bibitem[24]{key-16}P. Nicolini, A. Smailagic and E. Spallucci, Phys.
Lett. B \textbf{632}, 547 (2006).

\bibitem[25]{key-21}C. Corda, S. Chakraborty and S. Saha, EJTP \textbf{12},
No. IYL15-34, 107 (2015).

\bibitem[26]{key-22}A. F. Ali et al, EPL \textbf{112}, 20005 (2015).

\bibitem[27]{key-23}D. V. Fursaev, Phys. Rev. D \textbf{51}, 5352
(1995). 

\bibitem[28]{key-17}Y. S. Myung, Yong-Wan Kim and Young-Jai Park,
J. High Energy Phys. 02, 012(2007). 

\bibitem[29]{key-18}R. Banerjee, B. R. Majhi and Saurav Samanta,
Phys. Rev. D \textbf{77}, 124035 (2008).

\bibitem[30]{key-25}S. W. Hawking, \emph{The Path Integral Approach
to Quantum Gravity}. In \emph{General Relativity: An Einstein Centenary
Survey}; S. W. Hawking, W. Israel, Eds. Cambridge University Press:
Cambridge, UK, 1979. 

\bibitem[31]{key-26}S. Chakraborty, S. Saha, C. Corda, Galaxies \textbf{3},
103 (2015).

\bibitem[32]{key-28}J. D. Bekenstein, in Prodeedings of the Eight
Marcel Grossmann Meeting, T. Piran and R. Ruffini, eds., pp. 92-111
(World Scientific Singapore 1999). 

\bibitem[33]{key-33}N. Bohr, Philos. Mag. 26 , 1 (1913).

\bibitem[34]{key-34}N. Bohr, Philos. Mag. 26 , 476 (1913).

\bibitem[35]{key-36}N. Bohr, Zeits. Phys. 2, 423 (1920). 

\bibitem[36]{key-37}J. D. Bekenstein, Nuovo Cim. Lett. 4, 737 (1972). 

\bibitem[37]{key-38}J. D. Bekenstein, Phys. Rev. D7, 2333 (1973). 

\bibitem[38]{key-39}S. Shankaranarayanan, Mod. Phys. Lett. A 23,
1975-1980 (2008). 

\bibitem[39]{key-40}Hao-Peng Yan, Wen-Biao Liu, Phys. Lett. B 759,
293 (2016).

\bibitem[40]{key-41}R. Banerjee and B. R. Majhi, Phys. Lett. B 674,
218 (2009).

\bibitem[41]{key-42}R. Banerjee and B. R. Majhi, Phys. Rev. D 79,
064024 (2009).

\bibitem[42]{key-43}R. Banerjee, B. R. Majhi, and E. C. Vagenas,
Phys. Lett. B 686, 279 (2010).

\bibitem[43]{key-44}R. Banerjee, B. R. Majhi, and S. K. Modak, Class.
Quant. Grav. 26, 085010 (2009).

\bibitem[44]{key-45}G. Gecim, Y. Sucu, Phy. Lett. B, 773, 391 (2017).

\bibitem[45]{key-46}M. A. Anacleto, F. A. Brito, E. Passos, Phys.
Lett. B, 749, 181 (2015). 

\bibitem[46]{key-47}M. Faizal, M. M. Khalil, Int. J. Mod. Phys. A,
30, 1550144 (2015). 

\bibitem[47]{key-48}A. Ovgün and K. Jusufi, Eur. Phys. Jour. Plus,
132, 298 (2017).

\bibitem[48]{key-49}A. Ovgün and K. Jusufi, Eur. Phys. J. Plus, 131,
177 (2016).

\bibitem[49]{key-50}I. Sakalli, A. Ovgun and K. Jusufi, Astrop. Sp.
Sci. 361, 330 (2016).
\end{thebibliography}
\end{document}